\documentclass[prd,12pt,aps,superscriptaddress,double-spaced,floatfix,showkeys,showpacs]{revtex4}
\usepackage{graphicx,amsmath,bbm,bm,array,subfigure}
\usepackage[linktocpage=true,colorlinks=true,linkcolor=blue,citecolor=blue]{hyperref}

\begin{document}

\title {Hadron resonance gas EoS and the fluidity of matter produced in HIC}
\author{Guruprasad Kadam }
\email{guruprasadkadam18@gmail.com}
\affiliation{Department of Physics, Shivaji University,
Kolhapur, Maharashtra - 416 004, India}
\author{Swapnali Pawar }
\email{swapna191286@gmail.com}
\affiliation{Department of Physics, The New College,
Kolhapur, Maharashtra - 416 004, India}

%\date{\today} 

\def\be{\begin{equation}}
\def\ee{\end{equation}}
\def\bearr{\begin{eqnarray}}
\def\eearr{\end{eqnarray}}
\def\zbf#1{{\bf {#1}}}
\def\bfm#1{\mbox{\boldmath $#1$}}
\def\hf{\frac{1}{2}}
\def\sl{\hspace{-0.15cm}/}
\def\omit#1{_{\!\rlap{$\scriptscriptstyle \backslash$}
{\scriptscriptstyle #1}}}
\def\vec#1{\mathchoice
        {\mbox{\boldmath $#1$}}
        {\mbox{\boldmath $#1$}}
        {\mbox{\boldmath $\scriptstyle #1$}}
        {\mbox{\boldmath $\scriptscriptstyle #1$}}
}

\begin{abstract}
   We study the equation of state (EoS) of  hot and dense hadron gas by incorporating the excluded volume corrections into the ideal hadron resonance gas (HRG) model. The total  hadron mass spectrum  of the model is the sum of discrete mass spectrum consisting all the experimentally known hadrons  and the exponentially rising continuous Hagedorn states. We confront the EoS of the model with lattice quantum chromodynamics (LQCD) results at finite baryon chemical potential. We find that this modified HRG model reproduce the LQCD results up to $T=160$ MeV at zero as well as  finite baryon chemical potential. We further estimate the  shear viscosity within ambit of this model in the context of heavy-ion collision experiments.
\end{abstract}


\maketitle

\section{Introduction}
The phase diagram of quantum chromodynamics (QCD) is under intense theoretical investigation especially in the context of heavy-ion collision (HIC) experiments where two heavy nuclei are accelerated to very high energies and then they collide to produce very hot and dense QCD matter. These experiments can probe part of QCD phase diagram where the most interesting non perturbative aspects of QCD lie.  The most reliable theoretical tool at zero baryon chemical potential is lattice quantum chromodynamics (LQCD). One of the most important prediction of LQCD simulation is that the phase transition from hadronic to quark-gluon-plasma (QGP) is an analytic crossover\cite{Aoki:2006we}.  However, at finite baryon chemical potential QCD has been plagued with so called sign problem and one has to resort on certain approximation schemes to fetch the reliable results\cite{Lombardo:2005gj}. One can adopt an alternative approach where the effective model of QCD  can be constructed which not only preserve certain important symmetries of QCD but they are tractable at finite chemical potential as well.  The hadron resonance gas model is the statistical model of QCD describing low temperature hadronic phase of quantum chromodynamics. The essential starting point of the model is  Dashen-Ma-Bernstein theorem\cite{Dashen:1969ep} which allows us to compute the partition function of the interacting system in terms of scattering matrix. Using this theorem it can be shown that if the dynamics of the system is dominated by narrow-resonance formation it behaves like non-interacting system\cite{Dashen:1974jw,Welke:1990za,Venugopalan:1992hy}. Thus the thermodynamics of interacting gas of hadrons through formation of resonances can be well approximated by the non-interacting gas of hadrons and resonances. HRG model has been very successful in describing the hadron multiplicities in HICs\cite{ BraunMunzinger:1995bp,Yen:1998pa,Becattini:2000jw,Cleymans:1992zc,BraunMunzinger:2001ip,Rafelski:2002ga,Andronic:2005yp,Chatterjee:2013yga,Chatterjee:2015fua,Dash:2018nnj}. Recently, the interacting HRG model with multicomponent hard-core repulsion has been successful in describing the heavy ion collision data with the unprecedented accuracy\cite{Bugaev:2018gsh,Sagun:2017eye,Bugaev:2013sfa,Bugaev:2012wp}.

One possible improvement in the HRG model is to include the exponentially rising Hagedorn density of states apart from the known hadrons and resonances included in the form of discrete mass spectrum. This exponential mass spectrum arises in the string picture\cite{  Fubini:1970xj,Bardakci:1970yd,Fubini:1969wp,Huang:1970iq,Cohen:2006qd} or the glueball picture of the hadrons\cite{Thorn:1980iv}. A typical form of the continuous mass spectrum is $\rho(m)\sim A_{1}m^{-a}e^{A_{2}m}$ which satisfies the statistical bootstrap condition\cite{Hagedorn:1965st,Frautschi:1971ij}. With proper choice of the parameters $A_{1},A_{2}$ and with $a>5/2$ all the experimentally found hadrons fit in this exponential mass spectrum\cite{Broniowski:2000bj,Broniowski:2004yh}. Finite temperature LQCD simulations provides strong evidence of the existence of Hagedorn states in hot and dense matter created in heavy-ion collision experiments. The mass spectrum of these states are found to be of the form  $\rho(m)\sim m^{-a}e^{m/T_{H}}$ ($T_H$ is the Hagedorn temperature)\cite{Borsanyi:2010cj}.

Another possible improvement in the ideal HRG model is to take in to account short range repulsive interactions between hadrons. There are many  different ways of incorporating the repulsive interactions in the ideal HRG model without spoiling the thermodynamical consistency of the model\cite{Rischke:1992rk,Kapusta:1982qd,Singh:1991np}. We shall consider excluded volume correction scheme of Ref.\cite{Rischke:1992rk} where the repulsive interactions are accounted through the excluded volume correction in the ideal gas partition function.  

 In the past few decades the LQCD results of the equation of state at zero baryon chemical potential has been analyzed within HRG model and its extensions\cite{Karsch:2003vd,Karsch:2003zq,Redlich:2004gp,Albright:2014gva,Andronic:2012ut,Kadam:2014cua}.  Recently, in Ref.\cite{Vovchenko:2014pka} authors observed that the hadron resonance gas model with discrete mass spectrum augmented with the continuous hagedorn mass spectrum is not sufficient to explain the recent lattice QCD results. Further the excluded volume corrections to the ideal HRG model also fails to do the same. But if both these physical effects, $viz.$, the Hagedorn states as well as excluded volume corrections, are included in the ideal HRG model then the model reproduces the lattice QCD data all the way up to $T=160$MeV. 
 
In the context of relativistic heavy-ion collision (RHIC) experiments shear viscosity coefficient govern the evolution of the nonequilibrium system towards equilibrium state. In the off-central nuclear collision the spatial anisotropy of the produced matter gets converted into momentum anisotropy  and the equilibration of this momentum anisotropy is governed by shear viscosity coefficient. The produced matter in the fireball after the collision, with quarks and gluons degrees of freedom, behaves like a strongly interacting liquid with very small shear viscosity. Assuming that this liquid of quarks and gluons is in thermal equilibrium, it expands due to pressure difference and cools and finally undergo a phase transition to hadronic degrees of freedom which  finally free stream to the detector. One of the successful description of such an evolution is through dissipative relativistic hydrodynamics ~\cite{Gale:2013da,Schenke:2011qd,Shen:2012vn,Kolb:2003dz,Teaney:2000cw,DelZanna:2013eua,Karpenko:2013wva,
Holopainen:2011hq,Jaiswal:2015mxa}, and transport simulations~\cite{Xu:2004mz,Bouras:2010hm,Bouras:2012mh,Fochler:2010wn,Wesp:2011yy,Uphoff:2012gb,Greif:2013bb,Danielewicz:1984ww}. 
 In the hydrodynamic description of the heavy-ion collision experiments finite but small shear viscosity ($\eta$) to entropy ($s$) ratio is necessary to explain the flow data~\cite{Gyulassy:2004zy,Csernai:2006zz}. The smallness of this ratio $\frac{\eta}{s}$ and its connection to the conjectured Kovtun--Son-Starinets (KSS) bound 
of $\frac{\eta}{s}=\frac{1}{4\pi}$ obtained using AdS/CFT correspondence \cite{Kovtun:2004de} has motivated many theoretical investigations of this ratio to understand and derive rigorously from a microscopic theory~\cite{Gavin:1985ph,Prakash:1993bt,Dobado:2003wr,Itakura:2007mx,Chen:2007xe,Dobado:2009ek,Demir:2008tr, Puglisi:2014pda,Thakur:2017hfc,Tawfik:2010mb,Kadam:2018jaj,Denicol:2013nua, Khvorostukhin:2010aj,Kadam:2015xsa,FernandezFraile:2009mi,Kadam:2017iaz}.

In this work we confront the equation of state of excluded volume HRG model which includes the discrete hadron states and continuum Hagedorn states in the density of states  at finite baryon chemical potential. We further attempt to make rough estimates of the shear viscosity coefficient within ambit of this extended HRG  model in the context of heavy-ion collision experiments.  Throughout the discussion we shall adopt Boltzmann approximation (i.e Boltzmann classical statistics) since it is rather excellent approximation in the region of QCD phase diagram in which we are interested in (i.e $T=100-160$ MeV).

We organize the paper as follows. In section II we briefly describe the thermodynamics of hadron resonance gas model. In section III we give brief derivation of shear viscosity coefficient for the multicomponent hadronic matter using relativistic Boltzmann equation in relaxation time approximation. In section IV we  present the results  and discuss the implications of these results in the context of relativistic heavy ion collision experiments. Finally we  summarize and conclude in section V. 
 
\section{ Hadron resonance gas model}
Thermodynamical properties of hadron resonance gas model can be deduced from the grand canonical partition function defined as
\be
\mathcal{Z}(V,T,\mu_{B})=\int dm [\rho_{b}(m)\: \text{ln}Z_{b}(m,V,T,\mu_{B})+\rho_{f}(m)\: \text{ln}Z_{f}(m,V,T,\mu_{B})]
\ee
where $\mu_{B}$ is the baryon chemical potential and $\rho_{b}$ and $\rho_{f}$ are the mass spectrum of the bosons and fermions respectively.  We assume that the hadron mass spectrum is a combination of discrete (HG) and continuous Hagedorn states (HS) states given by
\be
\rho(m)=\rho_{HG}(m)+\rho_{HS}(m)
\ee
where 
\be
\rho_{HG}=\sum_{a}^{\Lambda}g_{a}\delta(m-m_{a})\theta(\Lambda-m)
\label{rhotot}
\ee

This discrete mass spectrum consists of all the experimentally known hadrons with cut-off $\Lambda$. One can set different cut-off values for baryons and mesons. The Hagedorn density of states is assumed to be 

\be
\rho_{HS}=Ae^{\frac{m}{T_{H}}}
\label{hs}
\ee
 where $A$ is constant and $T_{H}$ is the Hagedorn temperature. One can physically interpret $T_H$ as QCD phase transition temperature. It is important to note that one cannot define baryon chemical potential for the Hadedorn states. Hagedorn states depend on the trend in the increase of the number of states as the hadron mass increases. Since this is not a thermodynamic statement, concepts such as the baryon density or chemical potential would not apply to Hagedorn states. Further the Hagedorn states are defined by the density of states defined by Eq. (\ref{hs}) satisfying bootstrap condition. Since the quark content of these states is unknown it is not legitimate to define baryon number and hence baryon chemical potential to these states. Thus for all practical purpose we set $\mu_{B}=0$ for Hagedorn states in our calculation. It is to be noted that the heavy Hagedorn states has finite width and large width of Hagedorn states is of great
importance to describe the lattice QCD thermodynamics  and to explain the chemical equilibrium of hadronic matter born from these states \cite{Bugaev:2008iu,Beitel:2014kza}. But we will work in narrow width approximation for Hagedorn states.
 
Repulsive interaction can be accounted in the ideal HRG model via excluded volume correction ($V-vN$) to the partition function. The pressure of the HRG model with the discrete mass spectrum and excluded volume correction (which we shall call EHRG) turns out to be\cite{Rischke:1992rk}
\be
P^{EV}(T,\mu_{B})=\sum_{a}P^{id}_{a}(T,\tilde\mu_{B})
\label{prexl1}
\ee
where $\tilde\mu_{B}=\mu_{B}-vP^{EV}(T,\mu_{B})$, $P^{id}$ is the ideal gas pressure and $v=4\frac{4}{3}\pi r_{h}^3$ is the excluded volume parameter of the hadron with hard-core radius $r_h$. Note that we assume uniform hard-core radius to all the hadrons. The excluded volume models can be extended to multicomponent gas having different hard-core radius\cite{Gorenstein:1999ce}. These models can further be extended to take into account the Lorentz contraction of hadrons due to their relativistic motion\cite{Bugaev:2000wz}. But we will neglect these effects for the sake of simplicity. In the Boltzmann approximation the contribution to the ideal gas pressure due to $a^{th}$ hadronic species is
\be
P^{id}_{a}(T,\mu_{B})=\frac{g_{a}}{2\pi^2}\:T^2\:m_{a}^{2}\:K_{2}\bigg(\frac{m_{a}}{T}\bigg)
\ee
where $g_a$ is the degeneracy factor and $K_{n}$ is the modified Bessel's function of second kind.
 In the Boltzmann approximation Eq. (\ref{prexl1}) simplifies to
\be
P^{EV}(T,\mu_{B})=\sum_{a}P^{id}_{a}(T,\mu_{B})\:\text{exp}\bigg(-\frac{vP^{EV}(T,\mu_{B})}{T}\bigg)
\label{prexl}
\ee
 In the case of continuum Hagedorn spectrum the sum in Eq. (\ref{prexl}) is replaced by the integration and the discrete mass spectrum given by delta function is replaced by Hagedorn mass spectrum given by Eq. (\ref{hs}).  The other thermodynamical quantities, $viz.$, energy density ($\varepsilon(T,\mu_{B})$), entropy density ($s(T,\mu_{B})$), number density ($n(T,\mu_{B})$) and speed of sound ($C_{s}^2(T,\mu_{B})$) can be obtained from the pressure by taking appropriate derivatives as per thermodynamical identities.

\section{Transport coefficients: A Relaxation time approximation}
In the relativistic kinetic theory the evolution of distribution function $f_{p}(\bf{x},t)$ is determined by the Boltzmann equation\cite{landauPK}
\be
p^{\mu}\partial_{\mu}f_{p}=\mathcal{C}[f_{p}]
\label{Boltzmann}
\ee
where, $P^{\mu}=(E_{p},\bf{P})$ and $\mathcal{C}$ is called the collision term. Solving this equation, which is in general an "integro-differential" equation, for $f_{p}$ is rather very difficult task if not impossible. So it is customary to resort to certain approximations so that solving Eq. (\ref{Boltzmann}) becomes feasible.  We assume that the system is only slightly away from the equilibrium, i.e
\be
f_{p}=f_{p}^{0}+\delta f_{p}
\ee
with $f_{p}^{0}\gg\delta f_{p}$.  $f_{p}^{0}$ is an equilibrium distribution function. If we further assume that the collisions bring the system towards equilibrium with the time scale $\sim \tau$, then the collision term can be approximated as
\be
\mathcal{C}[f_{p}]\simeq -\frac{p^{\mu}u_{\mu}}{\tau(E_{p})}\delta f_{p}
\ee

In this so called relaxation time approximation the Boltzmann equation (\ref{Boltzmann}) becomes

\be
p^{\mu}\partial_{\mu}f_{p}^{0}=-\frac{p^{\mu}u_{\mu}}{\tau}\delta f_{p}
\label{RTA}
\ee

The equilibrium distribution function $f_{p}^{0}$ in the Boltzmann approximation  is given by
\be
f_{p}^{0}=\text{exp}\bigg\{-\frac{(E_{p}-\vec p.\vec u-\mu_{B})}{T}\bigg\}
\label{distribution}
\ee
where $\vec u$ is the fluid velocity. 

In the theory of fluid dynamics shear ($\eta$) and bulk ($\zeta$) viscosities enters as a coefficients of space-space component of the energy-momentum tensor away from equilibrium as
\be
T^{\mu\nu}=T_{0}^{\mu\nu}+T_{dissi}^{\mu\nu}
\ee
where $T_{0}^{\mu\nu}$ is the ideal part of stress tensor.

In the local Lorentz frame dissipative part of stress energy tensor can be written as
\be
T_{dissi}^{ij}=-\eta\bigg(\frac{\partial u^{i}}{\partial x^{j}}+\frac{\partial u^{j}}{\partial x^{i}}\bigg)-(\zeta-\frac{2}{3}\eta)\frac{\partial u^{i}}{\partial x^{j}}\delta^{ij}
\label{dissi}
\ee

In the kinetic theory $T^{\mu\nu}$ is defined as
\be
T^{\mu\nu}=\int d\Gamma\frac{p^{\mu}p^{\nu}}{E_{p}}f_{p}=\int d\Gamma\:\frac{p^{\mu}p^{\nu}}{E_{p}}(f_{p}^{0}+\delta f_{p})
\ee
where $d\Gamma=\frac{gd^{3}p}{(2\pi)^{3}}$, $g$ is the degeneracy. The space-space component of above equation is
\be
T^{ij}=T_{0}^{ij}+T^{ij}_{dissi}
\ee
where
\be
T_{dissi}^{ij}=\int d\Gamma\: p^{i}p^{j}\delta f_{p}
\label{dissi1}
\ee

In the local rest frame Eq. (\ref{RTA}) simplifies to
\be
\delta f_{p}=-\tau(E_{p})\bigg(\frac{\partial f_{p}^{0}}{\partial t}+ v_{p}^{i}\frac{\partial f_{p}^{0}}{\partial x^{i}} \bigg)
\label{deltadistr}
\ee
Assuming steady flow of the form $u^{i}=(u_{x}(y),0,0)$ and space-time independent temperature, Eq. (\ref{dissi}) simplifies to

\be
T^{xy}_{dissi}=-\eta\partial u_{x}/\partial y
\label{shear1}
\ee

From Eq. (\ref{dissi1}) and Eq. (\ref{deltadistr}) we get (using Eq.(\ref{distribution}) with $\mu_{B}=0$) 
\be
T^{xy}_{dissi}=\bigg\{-\frac{1}{T}\int d\Gamma\:\tau(E_{p})\bigg(\frac{p_{x}p_{y}}{E_{p}}\bigg)^{2}f_{p}^{0}\bigg\}\frac{\partial u_{x}}{\partial y}
\label{shear2}
\ee

Comparing the coefficients of gradients in Eqs. (\ref{shear1}) and (\ref{shear2}) we get the coefficient of shear viscosity
\be
\eta=\frac{1}{15T}\int d\Gamma\:\tau(E_{p})\frac{p^{4}}{E_{p}^{2}}f_{p}^{0}
\ee

% Bulk viscosity is related to the dissipation in the system when it is uniformly compressed.  Taking trace of Eq. (\ref{dissi}) we get
% \be
% (T_{dissi})^{i}_{i}=-3\zeta \frac{\partial u^{i}}{\partial x^{i}}
% \label{bulkdissi1}
% \ee
% Also from Eq. (\ref{dissi1}) and Eq. (\ref{deltadistr}) we get
% \be
% (T_{dissi})^{i}_{i}=-\int d\Gamma\:\tau(E_{p})\frac{p^{2}}{E_{p}}\bigg(\frac{\partial f_{p}^{0}}{\partial t}+ v_{p}^{i}\frac{\partial f_{p}^{0}}{\partial x^{i}} \bigg)
% \label{bulkdissi2}
% \ee
% Using energy momentum conservation law $\partial_{\mu}T^{\mu\nu}=0$, together with Eq. (\ref{bulkdissi1}) and Eq. (\ref{bulkdissi2}) we get the coefficient of bulk viscosity
% \be
% \zeta=\frac{1}{T}\int d\Gamma\:\tau(E_{p}) f^{0}_{p}\bigg[E_{p}v_{n_{B}}^{2}-\frac{p^{2}}{3E_{p}}\bigg]^{2}
% \ee
% where $v_{n_{B}}^{2}=\frac{\partial P}{\partial\varepsilon}|_{n_{B}}$ is the speed of sound at constant baryon density.

Thus for multicomponent hadron gas at finite chemical potential shear viscosity coefficient is\cite{Gavin:1985ph}
\be
\eta=\frac{1}{15T}\sum_{a}\int d\Gamma_{a}\:\frac{{p}^{4}}{E_{a}^{2}}{\tau}_{a}(E_{a})f_{p}^{0}
\label{shearmulti}
\ee

% \be
% \zeta=\frac{1}{T}\sum_{a}\int d\Gamma_{a}\:\tau_{a}(E_{a}) f^{0}_{a}\bigg[E_{a}v_{n_{B}}^{2}+\bigg(\frac{\partial P}{\partial n_{B}}\bigg)_{\varepsilon} -\frac{ p^{2}}{3E_{a}}\bigg]^{2}
% \label{blkmulti}
% \ee

The relaxation time is in general energy dependent. But for the simplicity we use averaged relaxation time ($\tilde \tau$) which is rather a good approximation as energy dependent relaxation time\cite{Moroz:2013haa}. 
Thus averaged partial relaxation time is defined by
\be
{\tilde\tau}_{a}^{-1}=\sum_{b}n_{b}\langle\sigma_{ab}v_{ab}\rangle
\label{relxaverage}
\ee
where $n_{b}=\int d\Gamma f_{b}^{0}$ is the number density of $b^{th}$ hadronic species and $\langle\sigma_{ab}v_{ab}\rangle$ is the thermal average of the cross section given by\cite{Gondolo:1990dk}
\be
\langle\sigma_{ab}v_{ab}\rangle=\frac{\sigma}{8Tm_{a}^{2}m_{b}^{2}K_{2}(\frac{m_{a}}{T})K_{2}(\frac{m_{b}}{T})}\int_{m_{a}+m_{b}}^{\infty}dS\:\frac{[S-(m_{a}-m_{b})^{2}]}{\surd S}[S-(m_{a}+m_{b})^{2}]K_{1}(\surd S/T)
\label{thermalave1}
\ee
where $S$ is a center of mass energy and $K_{n}$ is the modified Bessel's function of second kind. Note that we have assumed  the cross section $\sigma$ to be constant for all the species in the system. 

The massive Hagedorn states cannot be described rigorously using the Boltzmann equation. Thus it is very difficult to compute the contribution of these massive and highly unstable hadrons to the shear viscosity of low temperature QCD matter. But since these states contribute significantly to the thermodynamics of hadronic matter close to QCD transition temperature their contribution cannot be ignored. In fact they decay so rapidly that it is legitimate to assume that their presence will affect the relaxation time of the system. In Ref.\cite{NoronhaHostler:2008ju} authors studied the effect of Hagedorn states on the shear viscosity of QCD matter near $T_c$. They assumed that the relaxation time $\tau$ is inversely related to the decay width through relation $\tau=1/\Gamma$. The decay width of Hagedorn states can be obtained from the linear fit to the decay widths of all the resonances in the particle data book. This prescription corresponds to the decay cross section and neglect collisional cross section for the momentum transport that might contribute the shear viscosity as per Boltzmann equation. Such approximation gives only rough estimate of the shear viscosity. Since we are also interested in the rough estimate of shear viscosity we assume that the Hagedorn states contribute to the shear viscosity through collisions with other hadrons. We further assume that these states are hard sphere particles of radius $r_h$. Thus the the thermodynamics of the Hagedorn states can be estimated using excluded volume HRG model. It can be shown that in the excluded volume approximation the shear viscosity of  Hagedorn gas is\cite{NoronhaHostler:2012ug} 
\be
\eta_{HS}=\frac{5}{64r_{h}^2}\sqrt{\frac{T}{\pi}}\:\frac{T}{2\pi^2n(T)}\int_{0}^{\infty}dm\: \rho_{HS}(m)\:m^{5/2}\:K_{5/2}\bigg(\frac{m}{T}\bigg)
\ee
where $n(T)$ is the number density of Hagedorn gas. Viscosity coefficients of pure Hagedorn fluid in the relaxation time approximation of the Boltzmann equation has also been studied in Ref.\cite{Tawfik:2010mb}. We again mention here that the heavy Hagedorn states has finite width but we are working in the narrow width approximation of these states.

\section{Results and discussion}

 We include all the hadrons and resonances up 2.225 GeV listed in Ref.\cite{Amsler:2008zzb}. More specifically we choose cut-off $\Lambda_{M}=2.011$ GeV for mesons and $\Lambda_{B}=2.225$ GeV for the baryons. Note here that in HRGM usually one has to include all hadrons and resonances up to $2.5-2.6$ GeV in order to get a good description of the experimental data. But the mass spectrum of hadrons with masses between 1 GeV and 2.2 GeV is approximately Hagedorn-like.  We set uniform hardcore radius $r_{h}=0.4$ fm to all the hadrons. Note here that the excluded volume corrections to ideal HRG model is consistent only if we choose uniform hard core radius to all the hadrons. To specify the Hagedorn mass spectrum we choose, $A=0.4$ GeV$^{-1}$. Finally we set the Hagedorn temperature $T_{H}=198$ MeV for $\mu_{B}=0$ and $T_{H}=188$ MeV for $\mu_{B}=300$ MeV. This specific choice is made to get the best fit with the LQCD data at zero as well as finite chemical potential.

  \begin{figure}[t]
\vspace{-0.4cm}
\begin{center}
\begin{tabular}{c c}
 \includegraphics[width=9cm,height=9cm]{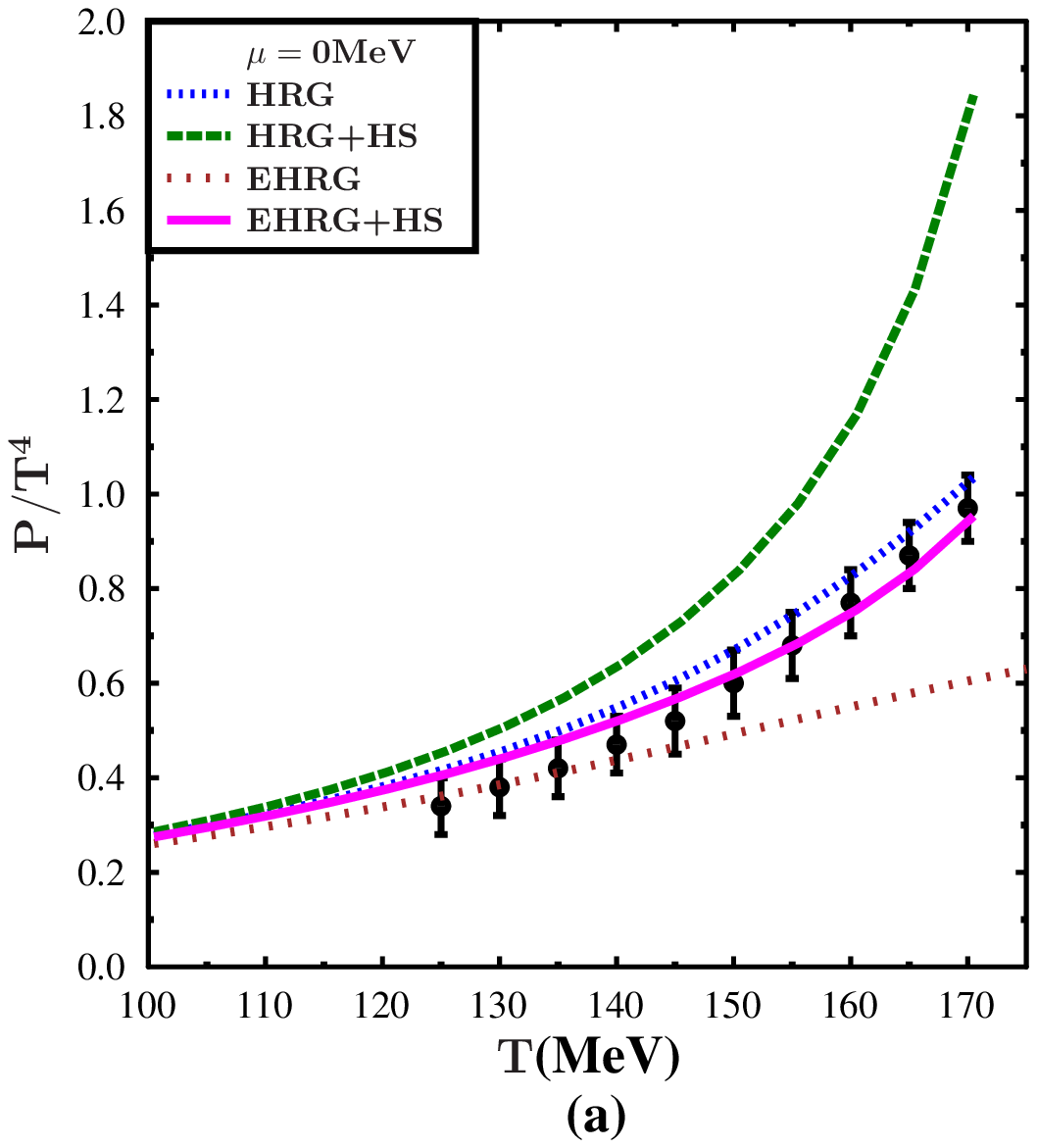}&
  \includegraphics[width=9cm,height=9cm]{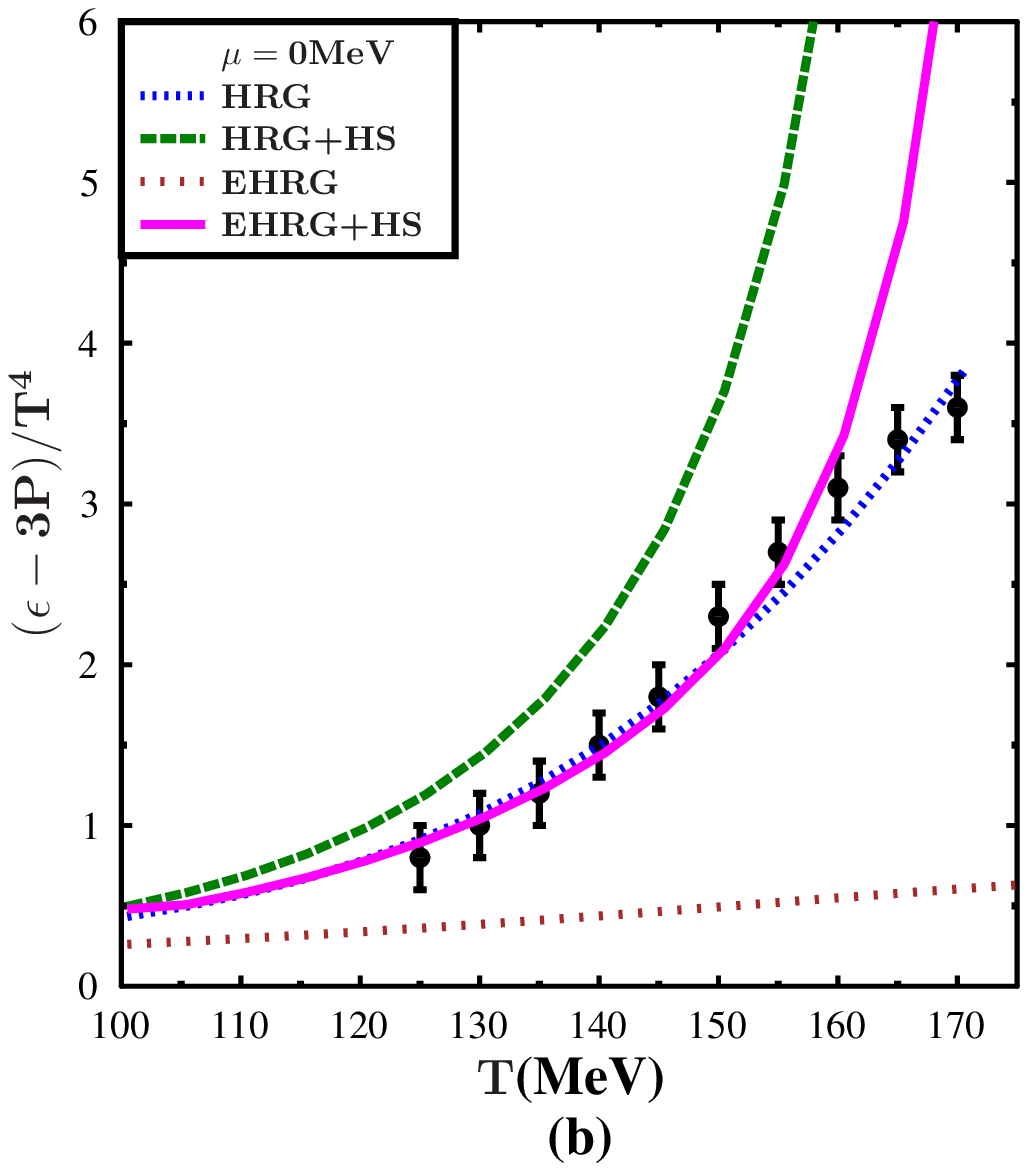}
  \end{tabular}
  \caption{ Thermodynamical functions, pressure (left panel) and trace anomaly (right panel) at zero chemical potential. The hard-core radius of all the hadrons has been set to the value $r_h=0.4 $fm. The Hagedorn mass spectrum is given by Eq. (\ref{hs}). The LQCD data has been taken from Ref. \cite{Borsanyi:2012cr}. } 
\label{thermu0}
  \end{center}
 \end{figure}
 
\begin{figure}[t]
\vspace{-0.4cm}
\begin{center}
\begin{tabular}{c c}
 \includegraphics[width=9cm,height=9cm]{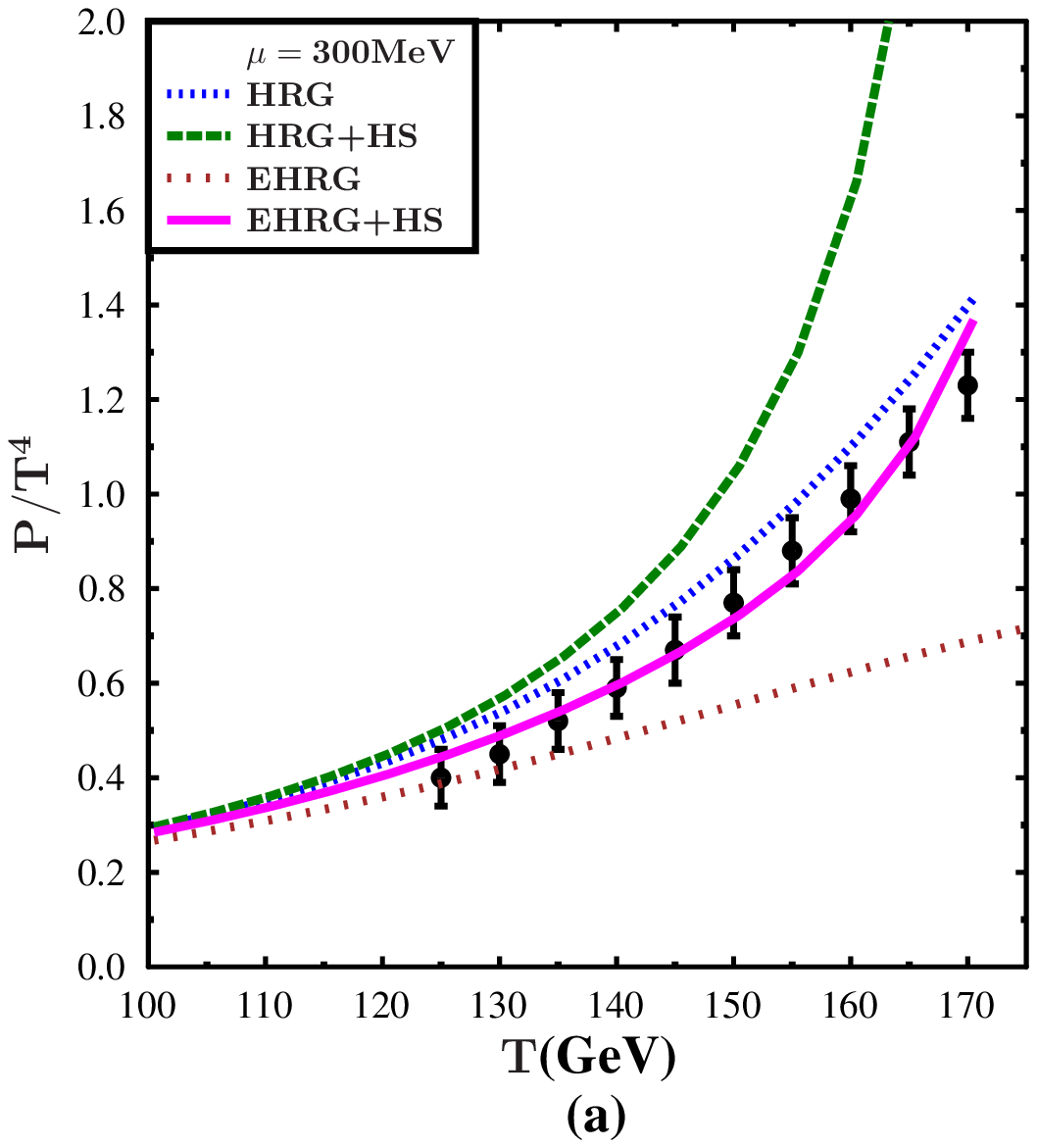}&
  \includegraphics[width=9cm,height=9cm]{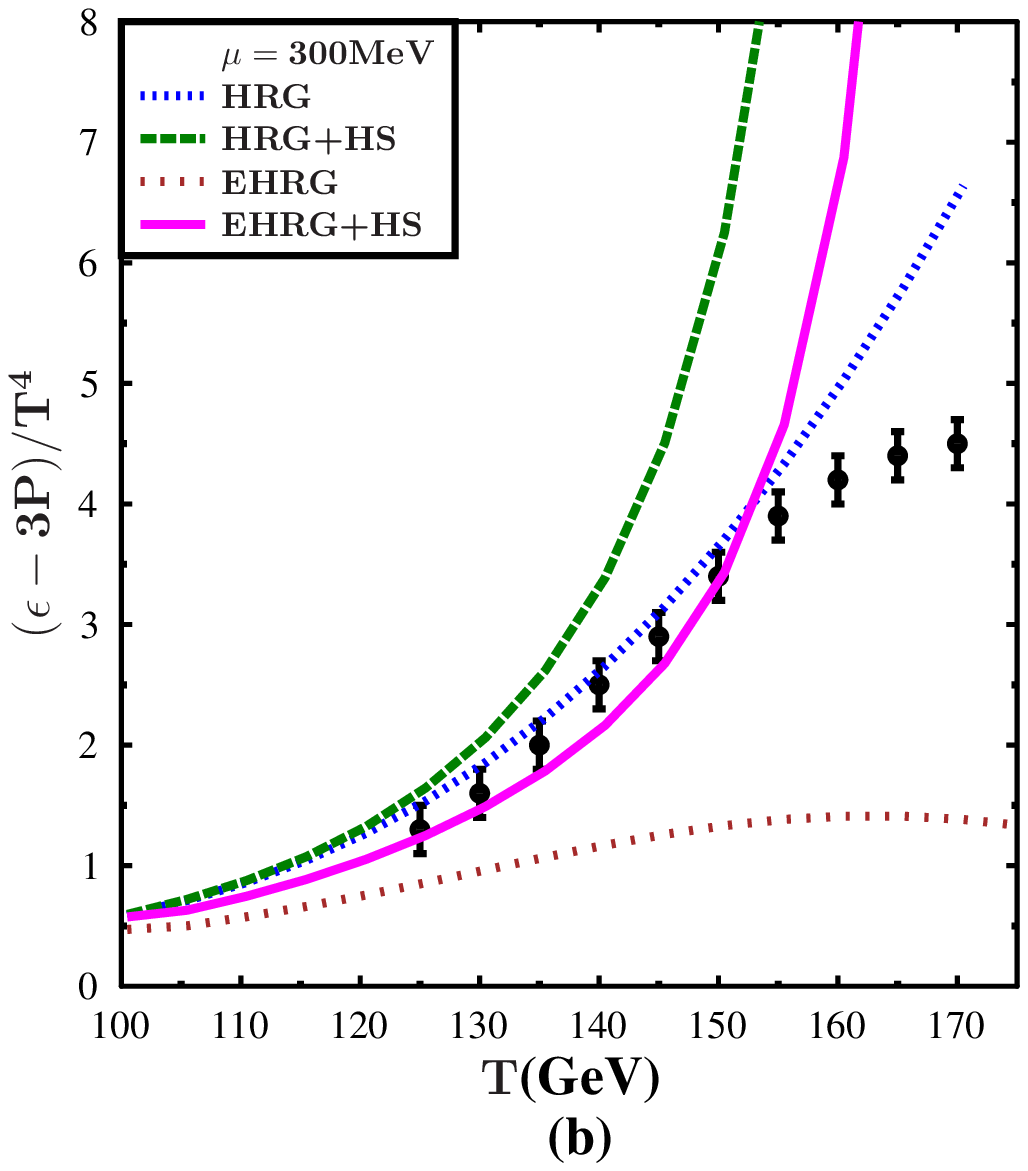}
  \end{tabular}
  \caption{ Thermodynamical functions, pressure (left panel) and trace anomaly (right panel) at finite chemical potential. The hard-core radius of all the hadrons has been set to the value $r_h=0.4 $fm. The LQCD data has been taken from the Ref.\cite{Borsanyi:2012cr}. } 
\label{thermu300}
  \end{center}
 \end{figure}

 Fig.(\ref{thermu0}) shows scaled pressure ($\frac{P}{T^4}$) and the scaled interaction measure ($I=\frac{\epsilon-3P}{T^4}$) at zero chemical potential. The black dots (with error bars) corresponds lattice QCD data taken from the Ref. \cite{Borsanyi:2012cr}. The blue curve corresponds to ideal HRG model with only discrete mass spectrum while the green dashed curve corresponds to ideal HRG with both discrete as well as continuous Hagedorn mass spectrum. Brown curve corresponds to  excluded volume HRG model with only experimentally known hadrons included. The solid magenta curve corresponds to excluded volume corrections to ideal HRG model in which both experimentally known hadrons as well as continuous Hagedorn states are included. We shall call HRG with excluded volume corrections and Hagedorn spectrum MEHRG (modified excluded volume HRG) for brevity. It can be  noted that HRG model alone cannot reproduce lattice data for the pressure as well as the interaction measure. The HRG estimates for the pressure are merely within the error bars. In case of interaction measure, while the LQCD predict the rapid rise till $T=160$ MeV, HRG estimates do not show such rapid rise. Similar points can be noted at finite chemical potential as shown in Fig.\ref{thermu300}. Further, the inclusion of Hagedorn mass spectrum or the excluded volume corrections to ideal HRG model do not improve the results.  But if excluded volume corrections are made to ideal HRG model with discrete as well as continuous Hagedorn mass spectrum then the resulting model (MEHRG) reproduce the LQCD results up to temperature $T=160$ MeV. This observation has already been made at zero chemical potential in Ref.\cite{Vovchenko:2014pka} where the authors confronted the LQCD data at $\mu_{B}=0$ with the HRG model with Hagedorn mass spectrum and excluded volume correction. We observe that the similar conclusion can be drawn at finite chemical potential as well. While the ideal HRG do not satisfactorily reproduce all the features of LQCD data, MEHRG reproduce the lattice data all the way up to $T=160$ MeV. It is to be noted in passing that at finite chemical potential QCD phase transition occurs at lower temperature than that at zero chemical potential. 
 
\begin{figure}[t]
\vspace{-0.4cm}
\begin{center}
 \includegraphics[width=9cm,height=9cm]{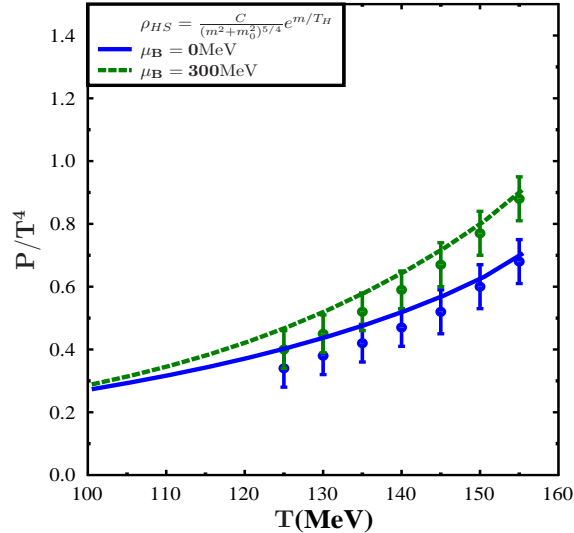}
  \caption{$P/T^4$ at zero as well as finite baryon chemical potential for the Hagedorn spectrum used in Ref.\cite{Vovchenko:2014pka}. The values of the parameters are also same as in Ref.\cite{Vovchenko:2014pka}.} 
\label{premu}
  \end{center}
 \end{figure}

 Fig.(\ref{premu}) shows $P/T^4$ estimated in MEHRG with the Hagedorn mass spectrum of the form $\rho_{HS}=\frac{C}{(m^2+m_{0}^2)e^{m/T_{H}}}$. The values of the parameters are the same as in Ref.\cite{Vovchenko:2014pka}. We note that our conclusion, i.e the ideal HRG need to be improved with excluded volume corrections as well as Hagedorn density of states, does not depend on the choice of Hagedorn mass spectrum.
 
  \begin{figure}[t]
\vspace{-0.4cm}
\begin{center}
 \includegraphics[width=9cm,height=9cm]{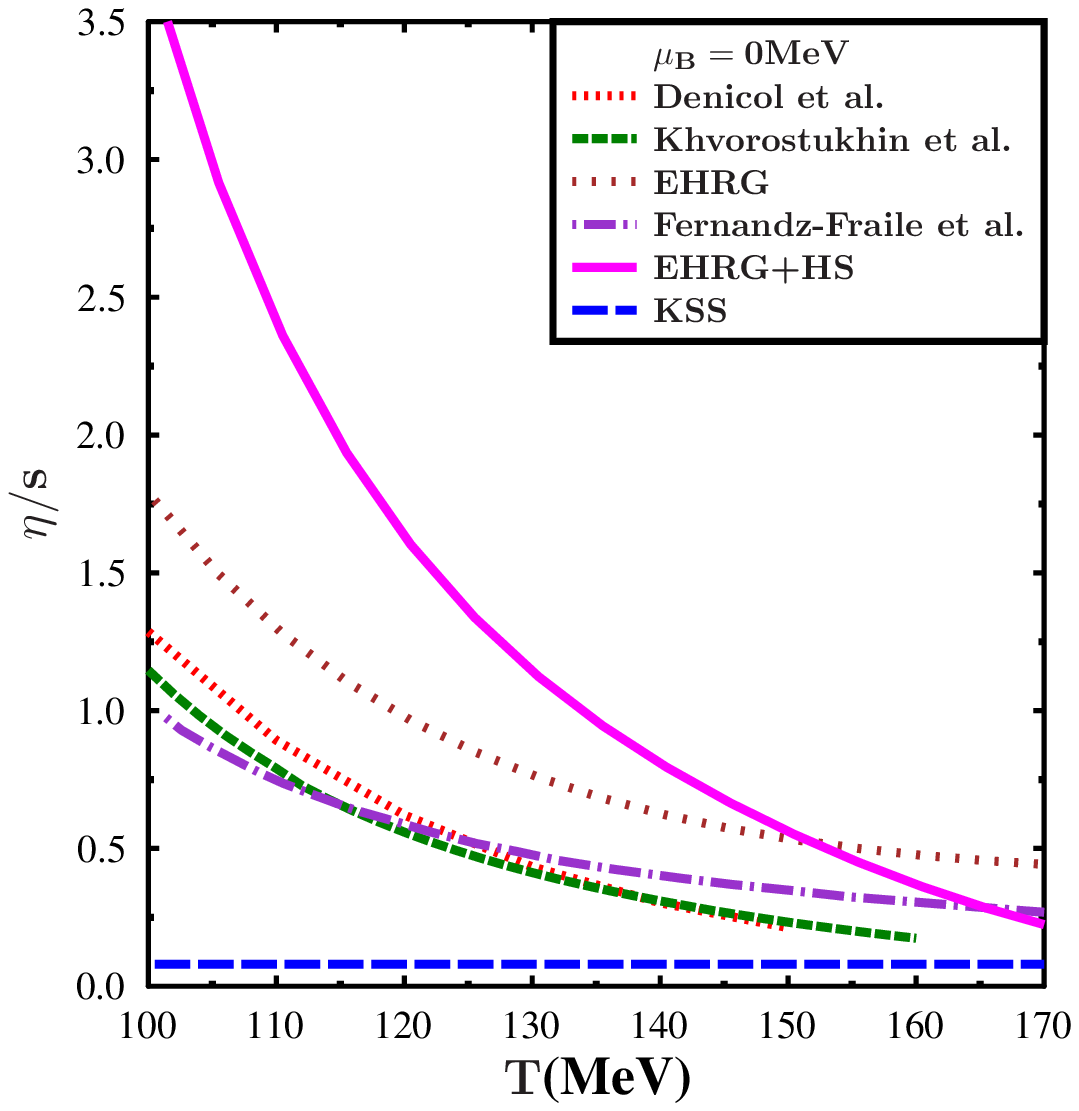}
  \caption{Comparison of the ratio $\eta/s$ estimated within our model i.e MEHRG (solid magenta curve) with various other methods. Red dashed curve corresponds to Chapman-Enscog method\cite{Denicol:2013nua} with constant cross sections. Dashed green curve corresponds to estimations made within SHMC model of the hadronic matter\cite{Khvorostukhin:2010aj}. Brown dashed curve corresponds to estimations made within EHRG model\cite{Kadam:2015xsa}. Dot-dashed orchid curve corresponds to the $\eta/s$ of meson gas estimated using chiral perturbation theory\cite{FernandezFraile:2009mi}.} 
\label{etabis2}
  \end{center}
 \end{figure}

 Fig. (\ref{etabis2}) shows comparison of the ratio $\eta/s$ estimated within our model with that of various other methods\cite{Denicol:2013nua, Khvorostukhin:2010aj,Kadam:2015xsa,FernandezFraile:2009mi}. Red dashed curve corresponds to Chapman-Enscog method with constant cross sections\cite{Denicol:2013nua}. Dashed green curve corresponds to relativistic Boltzmann equation in relaxation time approximation. The thermodynamical quantities in this model has been estimated  using scaled hadron masses and coupling (SHMC) model\cite{Khvorostukhin:2010aj}. Brown dashed curve corresponds to estimations made using relativistic Boltzmann equation in RTA. The thermodynamical quantities are estimated within EHRG model\cite{Kadam:2015xsa}. Dot-dashed orchid curve corresponds to the $\eta/s$ of meson gas estimated using chiral perturbation theory\cite{FernandezFraile:2009mi}. While the ratio $\eta/s$ in our model is relatively large at low temperature as compared to other models it rapidly falls and approaches closer to the Kovtun-Son-Starinets (KSS) bound\cite{Kovtun:2004de}, $\frac{\eta}{s}=\frac{1}{4\pi}$ at high temperature. This rapid fall may be attributed to the rapidly rising entropy density due to Hagedorn states which are absent in the other models.

\begin{figure}[t]
\vspace{-0.4cm}
\begin{center}
\begin{tabular}{c c}
 \includegraphics[width=9cm,height=9cm]{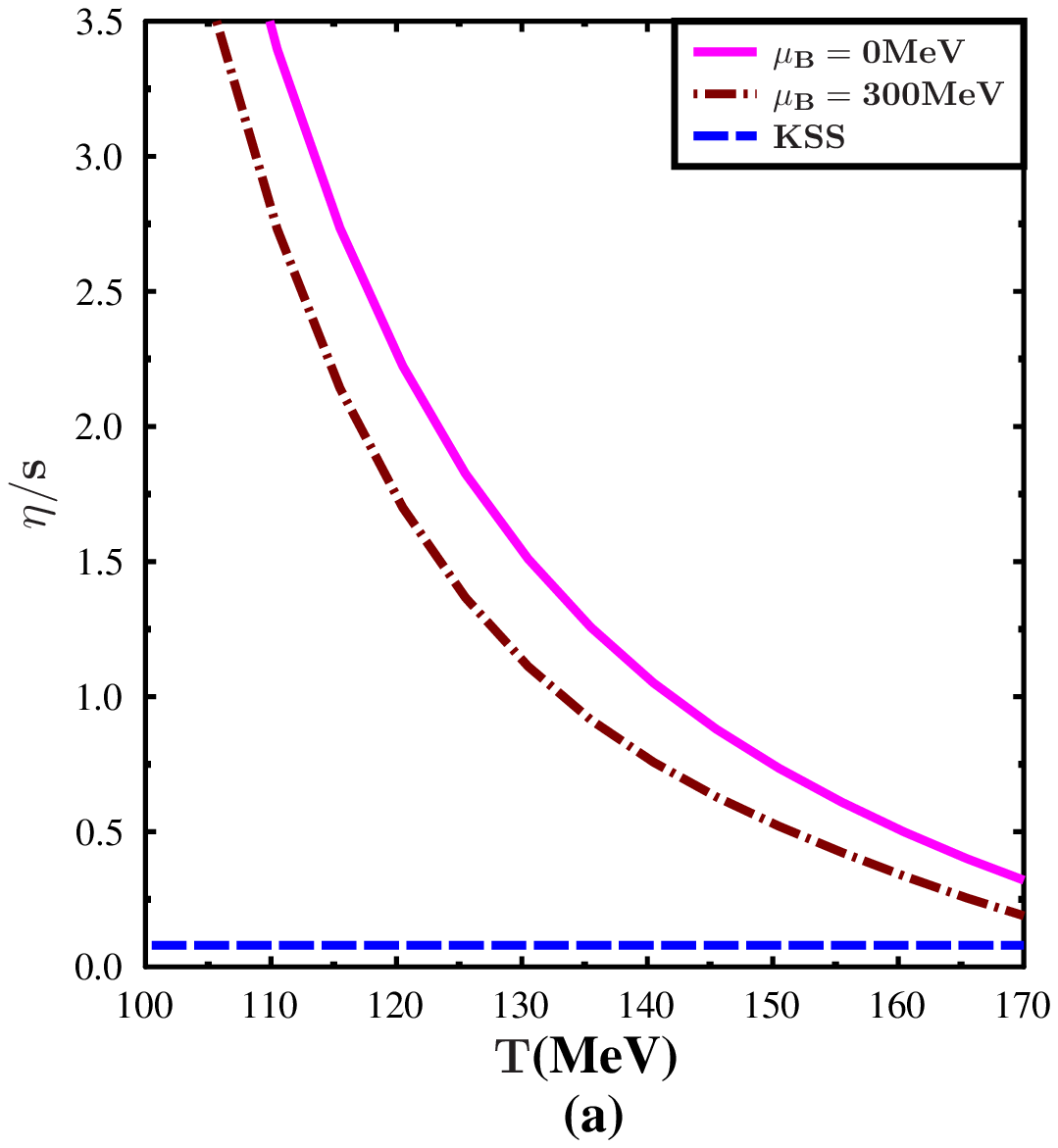}&
  \includegraphics[width=9cm,height=9cm]{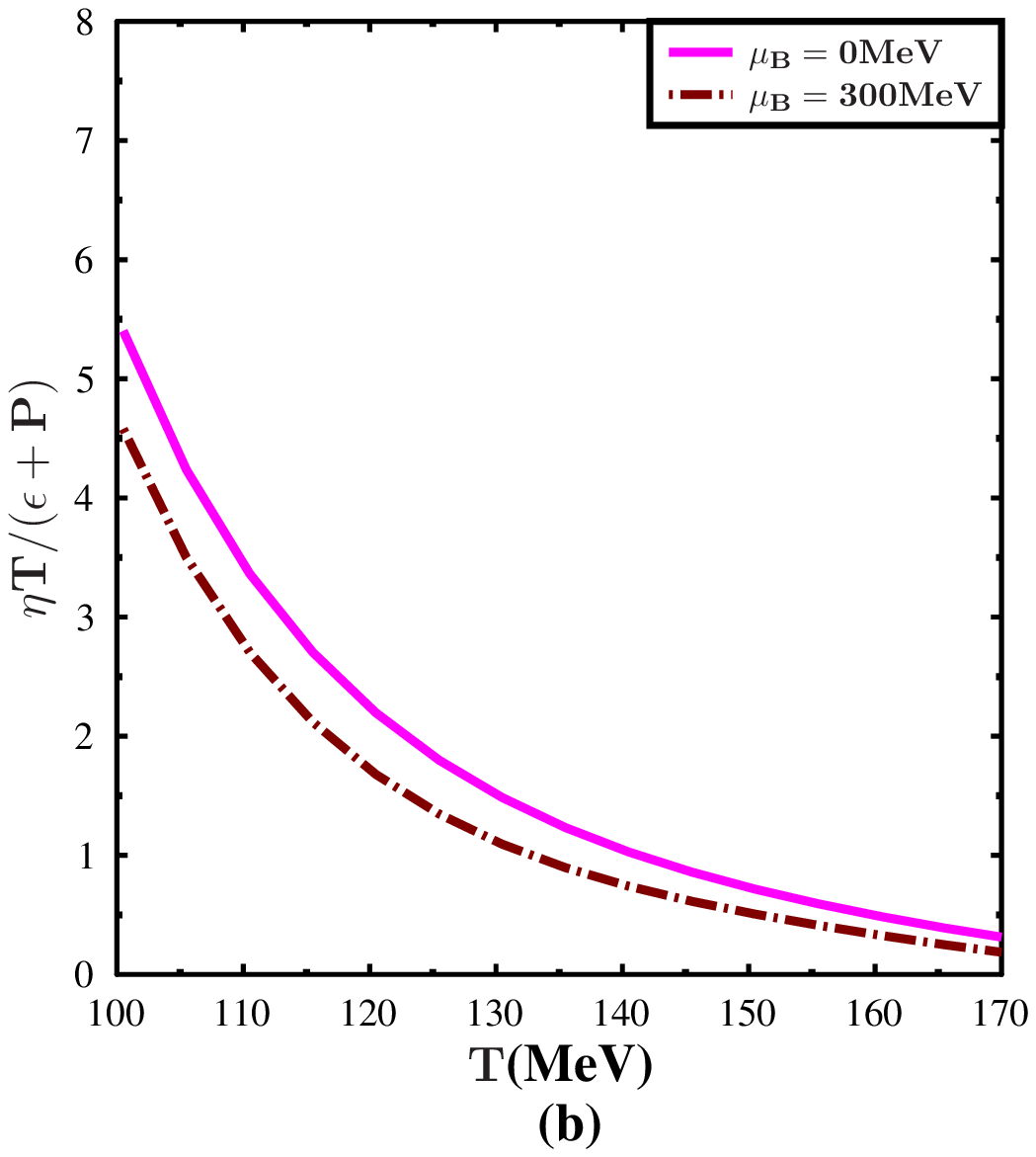}
  \end{tabular}
  \caption{ Left panel shows $\eta/s$ estimated within MEHRG model at $\mu_{B}=0$ and 300 MeV. The blue dashed curve corresponds to KSS bound, $\frac{\eta}{s}=\frac{1}{4\pi}$. Right panel shows the fluidity measure $\frac{\eta T}{(\epsilon+P)}$ estimated within MEHRG at $\mu_{B}=0$ and 300 MeV.} 
\label{etabis1}
  \end{center}
 \end{figure}
 
 Fig. (\ref{etabis1}a) shows ratio of shear viscosity to entropy density at two different chemical potentials. The ratio $\eta/s$ decreases with increase in temperature and approaches KSS bound at high temperature. The rapid fall in $\eta/s$ estimated within MEHRG can again be attributed to the rapid rise in the entropy density due to exponetially rising Hagedorn states. At finite $\mu_{B}$ this ratio approaches KSS bound more closely as compared to zero chemical potential case.
 
 It has been argued in Ref.\cite{Liao:2009gb} that at finite chemical potential correct fluidity measure is not $\eta/s$ but the quantity $\frac{\eta T}{(\epsilon+P)}$. At zero chemical potential basic thermodynamical identity implies that two fluidity measures $\eta/s$ and $\frac{\eta T}{(\epsilon+P)}$ are the same. However, at finite chemical potential they may differ. Fig. (\ref{etabis1}b) shows the fluidity measure $\frac{\eta T}{(\epsilon+P)}$ at zero as well as at finite chemical potential. It can be noted that $\frac{\eta T}{(\epsilon+P)}$ shows behavior similar to that of $\eta/s$.

\begin{figure}[t]
\vspace{-0.4cm}
\begin{center}
\begin{tabular}{c c}
 \includegraphics[width=9cm,height=9cm]{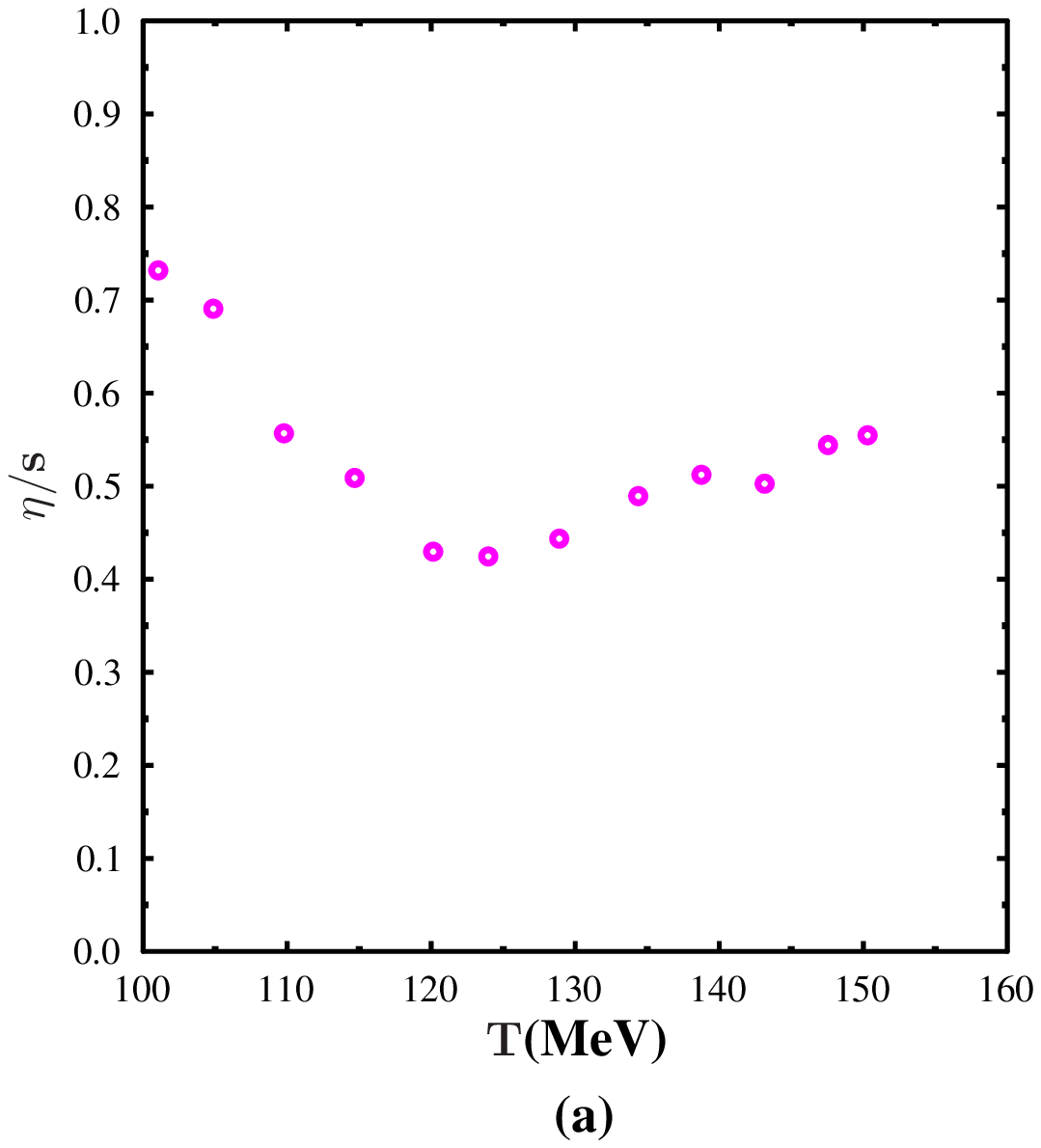}&
  \includegraphics[width=9cm,height=9cm]{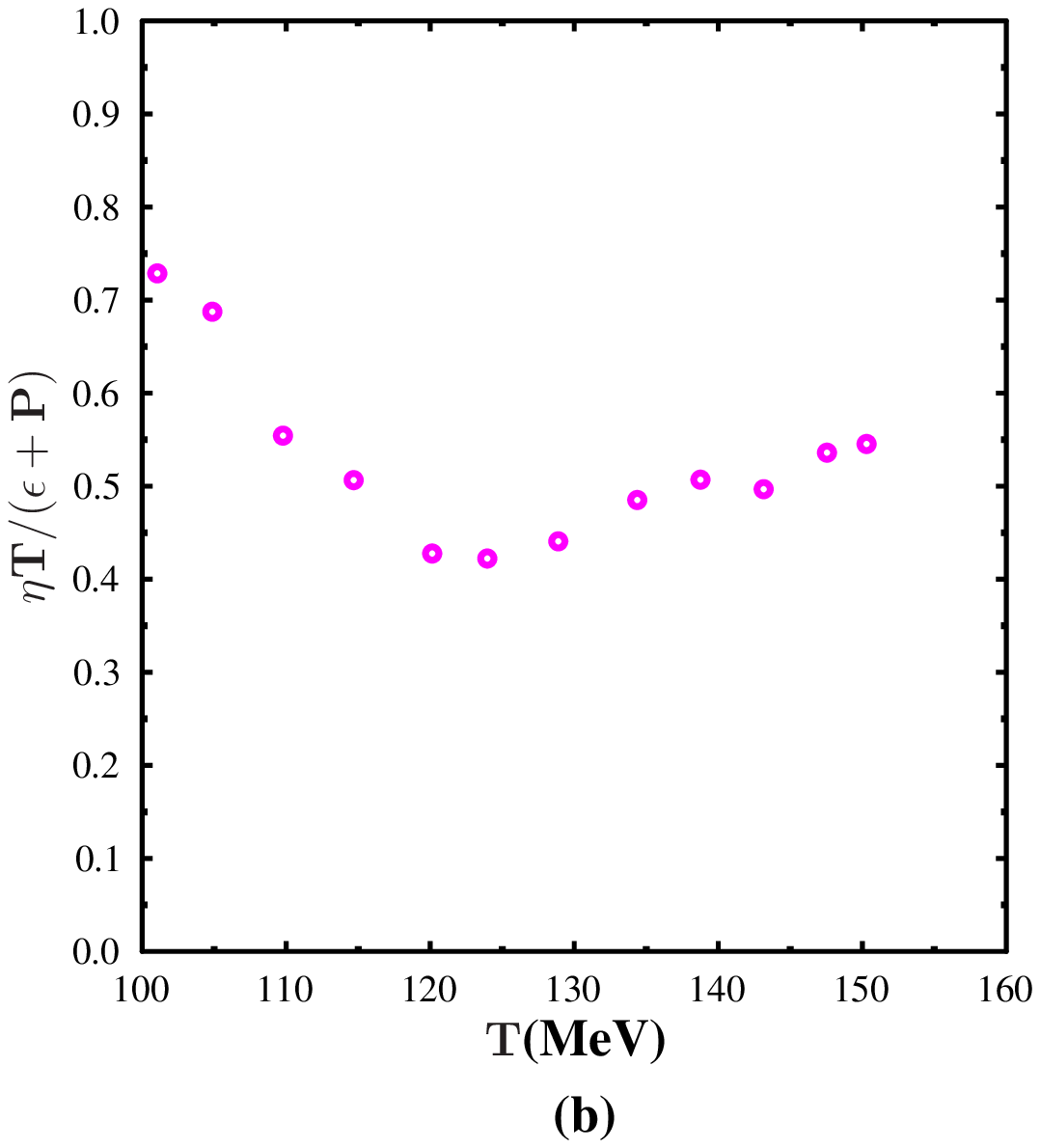}
  \end{tabular}
  \caption{ Fluidity measures along the chemical freeze-out line (Eqs. (\ref{freezT}) and (\ref{freezemu}) ). The freeze-out curve parametrization is taken from Ref.\cite{Bugaev:2013jza}.} 
\label{snn}
  \end{center}
 \end{figure}

 One can make estimations of $\eta/s$ in the context of heavy ion collision experiments by finding the beam energy ($\sqrt{S}$) dependence of the temperature and chemical potential. This is extracted from a statistical thermal model description of the particle yield at various $\sqrt{S}$ \cite{Cleymans:2005xv,Bugaev:2013jza,Tawfik:2012si,Tawfik:2013dba,Tawfik:2013eua,Tawfik:2016sqd,Tawfik:2018ahq}. The freeze out curve $T(\mu_{B})$ is parametrized by\cite{Bugaev:2013jza} 
 
 \be
 T(\sqrt{S_{NN}})=c_{+}(T_{10}+T_{20}\sqrt{S_{NN}})+c_{-}\bigg(T_{0}^{\text{lim}}+\frac{T_{30}}{\sqrt{S_{NN}}}\bigg)
 \label{freezT}
 \ee
 \be
 \mu(\sqrt{S_{NN}})=\frac{a_{0}}{1+b_{0}\sqrt{S_{NN}}}
 \label{freezemu}
 \ee
 where $T_{1O} = −34.4$ MeV, $T_{2O} = 30.9$ MeV$/$GeV, $T_{3O} = −176.8$ GeV MeV, $T_{0}^{\text{lim}} = 161.5$ MeV, $a_{0}=1481.6$ MeV and $b_{0}=0.365$ GeV$^{-1}$. The functions $c_{+}$ and $c_{-}$ smoothly connects the different behaviors of $\sqrt{S_{NN}}$. Fig.(\ref{snn}) shows fluidity measures, $\eta/s$ and $\frac{\eta T}{(\epsilon+P)}$ along chemical freeze-out line. It can be noted that the fluidity measures decreases as centre of mass energy increases, attains minimum, and then it does not vary much as $\sqrt{S}$ increases. This indicates that the fluid behavior of hadronic matter does not change much for the wide range of higher values of collision energies. Thus the matter produced in  heavy-ion collision experiments with wide range of high collision energies can exhibit substantial elliptic flow.  
 
 \newpage
 \section{Summary and conclusion}
 In this work we confronted the HRG equation of state with the LQCD results at finite baryon chemical potential. We noted that the ideal HRG model along with its extended forms, $viz.$, HRG with excluded volume corrections and HRG with discrete as well as continuous Hagedorn states cannot explain the lattice data simultaneously at zero as well as finite chemical potential. We found that the LQCD data can be accurately reproduced up to $T=160$ MeV at zero as well as finite baryon chemical potential if both the excluded volume corrections as well as Hagedorn states are simultaneously included in ideal HRG model. We noted that while the excluded volume corrections suppress the thermodynamical quantities as compared to ideal HRG results, the Hagedorn states provide necessary rapid rise in the trace anomaly observed in LQCD simulation results. With these observations we conclude that the unobserved heavy Hagedorn states plays important role in the thermodynamics of hadronic matter especially near QCD transition temperature.   
 
 We also estimated the shear viscosity coefficient within ambit of HRG model augmented with excluded volume corrections and Hagedorn states. We found that the behavior of the fluidity measures $\eta/s$ and $\eta T/(\epsilon+P)$  are in agreement with the existing results. We noted that both the fluidity measures falls rapidly at high temperature. This fall may be attributed to the the rapid rise in entropy density, $s$ and the quantity $(\epsilon+P)/T$ which are thermodynamically related to each other. We further noted that the fluidity measures do not change much at at high collision energies. This indicates that the matter produced in heavy-ion collision experiments with wide range of higher collision  energies can exhibit substantial elliptic flow.  

\newpage 
\section*{ACKNOWLEDGMENTS}
GK is financially supported by the DST-INSPIRE faculty award under the Grant No. DST/INSPIRE/04/2017/002293.


\begin{thebibliography}{99}

  \bibitem{Aoki:2006we} 
  Y.~Aoki, G.~Endrodi, Z.~Fodor, S.~D.~Katz and K.~K.~Szabo,
  %``The Order of the quantum chromodynamics transition predicted by the standard model of particle physics,''
  Nature {\bf 443}, 675 (2006)
  doi:10.1038/nature05120
  [hep-lat/0611014].
  
  \bibitem{Lombardo:2005gj} 
  M.~P.~Lombardo,
  %``Lattice QCD at finite temperature and density,''
  Mod.\ Phys.\ Lett.\ A {\bf 22}, 457 (2007)
  doi:10.1142/S0217732307023055
  [hep-lat/0509180].

  \bibitem{Dashen:1969ep}                                         
  R.~Dashen, S.~K.~Ma and H.~J.~Bernstein,
  %``S Matrix formulation of statistical mechanics,''
  Phys.\ Rev.\  {\bf 187}, 345 (1969).
  
  \bibitem{Dashen:1974jw} 
  R.~F.~Dashen and R.~Rajaraman,
  %``Narrow Resonances in Statistical Mechanics,''
  Phys.\ Rev.\ D {\bf 10}, 694 (1974).
  doi:10.1103/PhysRevD.10.694
  
  \bibitem{Welke:1990za} 
  G.~M.~Welke, R.~Venugopalan and M.~Prakash,
  %``The Speed of sound in an interacting pion gas,''
  Phys.\ Lett.\ B {\bf 245}, no. 2, 137 (1990).
  doi:10.1016/0370-2693(90)90123-N
   
   
   
  \bibitem{Venugopalan:1992hy} 
  R.~Venugopalan and M.~Prakash,
  %``Thermal properties of interacting hadrons,''
  Nucl.\ Phys.\ A {\bf 546}, 718 (1992).
  doi:10.1016/0375-9474(92)90005-5
  
  
  
  
  
  \bibitem{BraunMunzinger:1995bp} 
  P.~Braun-Munzinger, J.~Stachel, J.~P.~Wessels and N.~Xu,
  %``Thermal and hadrochemical equilibration in nucleus-nucleus collisions at the SPS,''
  Phys.\ Lett.\ B {\bf 365}, 1 (1996)
  doi:10.1016/0370-2693(95)01258-3
  [nucl-th/9508020].
  
  \bibitem{Yen:1998pa} 
  G.~D.~Yen and M.~I.~Gorenstein,
  %``The Analysis of particle multiplicities in Pb + Pb collisions at CERN SPS within hadron gas models,''
  Phys.\ Rev.\ C {\bf 59}, 2788 (1999)
  doi:10.1103/PhysRevC.59.2788
  [nucl-th/9808012].
  
  \bibitem{Becattini:2000jw} 
  F.~Becattini, J.~Cleymans, A.~Keranen, E.~Suhonen and K.~Redlich,
  %``Features of particle multiplicities and strangeness production in central heavy ion collisions between 1.7A-GeV/c and 158A-GeV/c,''
  Phys.\ Rev.\ C {\bf 64}, 024901 (2001)
  doi:10.1103/PhysRevC.64.024901
  [hep-ph/0002267].
  
  \bibitem{Cleymans:1992zc} 
  J.~Cleymans and H.~Satz,
  %``Thermal hadron production in high-energy heavy ion collisions,''
  Z.\ Phys.\ C {\bf 57}, 135 (1993)
  doi:10.1007/BF01555746
  [hep-ph/9207204].
  
  
  \bibitem{BraunMunzinger:2001ip} 
  P.~Braun-Munzinger, D.~Magestro, K.~Redlich and J.~Stachel,
  %``Hadron production in Au - Au collisions at RHIC,''
  Phys.\ Lett.\ B {\bf 518}, 41 (2001)
  doi:10.1016/S0370-2693(01)01069-3
  [hep-ph/0105229].
  
                
  
  \bibitem{Rafelski:2002ga} 
  J.~Rafelski and J.~Letessier,
  %``Testing limits of statistical hadronization,''
  Nucl.\ Phys.\ A {\bf 715}, 98 (2003)
  doi:10.1016/S0375-9474(02)01418-5
  [nucl-th/0209084].
  
  \bibitem{Andronic:2005yp} 
  A.~Andronic, P.~Braun-Munzinger and J.~Stachel,
  %``Hadron production in central nucleus-nucleus collisions at chemical freeze-out,''
  Nucl.\ Phys.\ A {\bf 772}, 167 (2006)
  doi:10.1016/j.nuclphysa.2006.03.012
  [nucl-th/0511071].
  
  
 
  
    \bibitem{Chatterjee:2013yga} 
  S.~Chatterjee, R.~M.~Godbole and S.~Gupta,
  %``Strange freezeout,''
  Phys.\ Lett.\ B {\bf 727}, 554 (2013)
  doi:10.1016/j.physletb.2013.11.008
  [arXiv:1306.2006 [nucl-th]].
  
  \bibitem{Chatterjee:2015fua} 
  S.~Chatterjee, S.~Das, L.~Kumar, D.~Mishra, B.~Mohanty, R.~Sahoo and N.~Sharma,
  %``Freeze-Out Parameters in Heavy-Ion Collisions at AGS, SPS, RHIC, and LHC Energies,''
  Adv.\ High Energy Phys.\  {\bf 2015}, 349013 (2015).
  doi:10.1155/2015/349013
  
  \bibitem{Dash:2018nnj} 
  A.~K.~Dash, R.~Singh, S.~Chatterjee, C.~Jena and B.~Mohanty,
  %``Role of system size in freeze-out conditions extracted from transverse momentum spectra of hadrons,''
  Phys.\ Rev.\ C {\bf 98}, no. 6, 064902 (2018)
  doi:10.1103/PhysRevC.98.064902
  [arXiv:1807.06829 [hep-ph]].
  
  \bibitem{Bugaev:2018gsh} 
  K.~A.~Bugaev, V.~V.~Sagun, A.~I.~Ivanytskyi, I.~P.~Yakimenko, E.~G.~Nikonov, A.~V.~Taranenko and G.~M.~Zinovjev,
  %``Going beyond the second virial coefficient in the hadron resonance gas model,''
  Nucl.\ Phys.\ A {\bf 970}, 133 (2018).
  doi:10.1016/j.nuclphysa.2017.11.008
  
  \bibitem{Sagun:2017eye} 
  V.~V.~Sagun {\it et al.},
  %``Hadron Resonance Gas Model with Induced Surface Tension,''
  Eur.\ Phys.\ J.\ A {\bf 54}, no. 6, 100 (2018)
  doi:10.1140/epja/i2018-12535-1
  [arXiv:1703.00049 [hep-ph]].
  
  
  \bibitem{Bugaev:2013sfa} 
  K.~A.~Bugaev, D.~R.~Oliinychenko, J.~Cleymans, A.~I.~Ivanytskyi, I.~N.~Mishustin, E.~G.~Nikonov and V.~V.~Sagun,
  %``Chemical Freeze-out of Strange Particles and Possible Root of Strangeness Suppression,''
  EPL {\bf 104}, no. 2, 22002 (2013)
  doi:10.1209/0295-5075/104/22002
  [arXiv:1308.3594 [hep-ph]].
  
  \bibitem{Bugaev:2012wp} 
  K.~A.~Bugaev, D.~R.~Oliinychenko, A.~S.~Sorin and G.~M.~Zinovjev,
  %``Simple Solution to the Strangeness Horn Description Puzzle,''
  Eur.\ Phys.\ J.\ A {\bf 49}, 30 (2013)
  doi:10.1140/epja/i2013-13030-y
  [arXiv:1208.5968 [hep-ph]].
  

  
 
  
 
  \bibitem{Fubini:1970xj} 
  S.~Fubini and G.~Veneziano,
  %``Duality in operator formalism,''
  Nuovo Cim.\ A {\bf 67}, 29 (1970).
  doi:10.1007/BF02728411
  

  
  \bibitem{Bardakci:1970yd} 
  K.~Bardakci and S.~Mandelstam,
  %``Analytic solution of the linear-trajectory bootstrap,''
  Phys.\ Rev.\  {\bf 184}, 1640 (1969).
  doi:10.1103/PhysRev.184.1640
  
  \bibitem{Fubini:1969wp} 
  S.~Fubini, D.~Gordon and G.~Veneziano,
  %``A general treatment of factorization in dual resonance models,''
  Phys.\ Lett.\  {\bf 29B}, 679 (1969).
  doi:10.1016/0370-2693(69)90109-9
  
  \bibitem{Huang:1970iq} 
  K.~Huang and S.~Weinberg,
  %``Ultimate temperature and the early universe,''
  Phys.\ Rev.\ Lett.\  {\bf 25}, 895 (1970).
  doi:10.1103/PhysRevLett.25.895
  
  \bibitem{Cohen:2006qd} 
  T.~D.~Cohen,
  %``QCD strings and the thermodynamics of the metastable phase of QCD at large N(c),''
  Phys.\ Lett.\ B {\bf 637}, 81 (2006)
  doi:10.1016/j.physletb.2006.04.035
  [hep-th/0602037].
  
  
  \bibitem{Thorn:1980iv} 
  C.~B.~Thorn,
  %``INFINITE N(c) QCD AT FINITE TEMPERATURE: IS THERE AN ULTIMATE TEMPERATURE?,''
  Phys.\ Lett.\  {\bf 99B}, 458 (1981).
  doi:10.1016/0370-2693(81)91179-5
  
  
  \bibitem{Hagedorn:1965st} 
  R.~Hagedorn,
  %``Statistical thermodynamics of strong interactions at high-energies,''
  Nuovo Cim.\ Suppl.\  {\bf 3}, 147 (1965).
  
  
  \bibitem{Frautschi:1971ij} 
  S.~C.~Frautschi,
  %``Statistical bootstrap model of hadrons,''
  Phys.\ Rev.\ D {\bf 3}, 2821 (1971).
  doi:10.1103/PhysRevD.3.2821
  
  
  \bibitem{Broniowski:2000bj} 
  W.~Broniowski and W.~Florkowski,
  %``Different Hagedorn temperatures for mesons and baryons from experimental mass spectra, compound hadrons, and combinatorial saturation,''
  Phys.\ Lett.\ B {\bf 490}, 223 (2000)
  doi:10.1016/S0370-2693(00)00992-8
  [hep-ph/0004104]
  
  
  \bibitem{Broniowski:2004yh} 
  W.~Broniowski, W.~Florkowski and L.~Y.~Glozman,
  %``Update of the Hagedorn mass spectrum,''
  Phys.\ Rev.\ D {\bf 70}, 117503 (2004)
  doi:10.1103/PhysRevD.70.117503
  [hep-ph/0407290].
  
  \bibitem{Borsanyi:2010cj} 
  S.~Borsanyi, G.~Endrodi, Z.~Fodor, A.~Jakovac, S.~D.~Katz, S.~Krieg, C.~Ratti and K.~K.~Szabo,
  %``The QCD equation of state with dynamical quarks,''
  JHEP {\bf 1011}, 077 (2010)
  doi:10.1007/JHEP11(2010)077
  [arXiv:1007.2580 [hep-lat]].
  
  
  
  
  
  \bibitem{landauPK}
   L. D. Landau and E. M. Lifshitz, Physical Kinetics (Pergamon, Oxford, 1981).
 
  \bibitem{Moroz:2013haa} 
  O.~Moroz,
  %``Shear and bulk viscosities of the hadron gas within relaxation time approximation and its test,''
  Ukr.\ J.\ Phys.\  {\bf 58}, 1127 (2013)
  [arXiv:1312.6429 [hep-ph]].
 
 
 \bibitem{Gondolo:1990dk} 
  P.~Gondolo and G.~Gelmini,
  %``Cosmic abundances of stable particles: Improved analysis,''
  Nucl.\ Phys.\ B {\bf 360}, 145 (1991).
  doi:10.1016/0550-3213(91)90438-4
  
  
  \bibitem{NoronhaHostler:2008ju} 
  J.~Noronha-Hostler, J.~Noronha and C.~Greiner,
  %``Transport Coefficients of Hadronic Matter near T(c),''
  Phys.\ Rev.\ Lett.\  {\bf 103}, 172302 (2009)
  doi:10.1103/PhysRevLett.103.172302
  [arXiv:0811.1571 [nucl-th]].
  
  
  \bibitem{NoronhaHostler:2012ug} 
  J.~Noronha-Hostler, J.~Noronha and C.~Greiner,
  %``Hadron Mass Spectrum and the Shear Viscosity to Entropy Density Ratio of Hot Hadronic Matter,''
  Phys.\ Rev.\ C {\bf 86}, 024913 (2012)
  doi:10.1103/PhysRevC.86.024913
  [arXiv:1206.5138 [nucl-th]].
  
  
  
  \bibitem{Amsler:2008zzb} 
  C.~Amsler {\it et al.} [Particle Data Group],
  %``Review of Particle Physics,''
  Phys.\ Lett.\ B {\bf 667}, 1 (2008).
  doi:10.1016/j.physletb.2008.07.018
  
  
  
  \bibitem{Borsanyi:2012cr} 
  S.~Borsanyi, G.~Endrodi, Z.~Fodor, S.~D.~Katz, S.~Krieg, C.~Ratti and K.~K.~Szabo,
  %``QCD equation of state at nonzero chemical potential: continuum results with physical quark masses at order $mu^2$,''
  JHEP {\bf 1208}, 053 (2012)
  doi:10.1007/JHEP08(2012)053
  [arXiv:1204.6710 [hep-lat]].
  
  
   
  \bibitem{Gorenstein:1999ce} 
  M.~I.~Gorenstein, A.~P.~Kostyuk and Y.~D.~Krivenko,
  %``Van der Waals excluded volume model of multicomponent hadron gas,''
  J.\ Phys.\ G {\bf 25}, L75 (1999)
  doi:10.1088/0954-3899/25/9/102
  [nucl-th/9906068].
  
  \bibitem{Bugaev:2000wz} 
  K.~A.~Bugaev, M.~I.~Gorenstein, H.~Stoecker and W.~Greiner,
  %``Van der Waals excluded volume model for Lorentz contracted rigid spheres,''
  Phys.\ Lett.\ B {\bf 485}, 121 (2000)
  doi:10.1016/S0370-2693(00)00690-0
  [nucl-th/0004061].
  
  
  
  
  
   
  \bibitem{Bugaev:2008iu} 
  K.~A.~Bugaev, V.~K.~Petrov and G.~M.~Zinovjev,
  %``Quark Gluon Bags as Reggeons,''
  Phys.\ Rev.\ C {\bf 79}, 054913 (2009)
  doi:10.1103/PhysRevC.79.054913
  [arXiv:0807.2391 [hep-ph]].
  
  \bibitem{Beitel:2014kza} 
  M.~Beitel, K.~Gallmeister and C.~Greiner,
  %``Thermalization of Hadrons via Hagedorn States,''
  Phys.\ Rev.\ C {\bf 90}, no. 4, 045203 (2014)
  doi:10.1103/PhysRevC.90.045203
  [arXiv:1402.1458 [hep-ph]].
  
  
  
  \bibitem{Rischke:1992rk} 
  D.~H.~Rischke, J.~Schaffner, M.~I.~Gorenstein, A.~Schaefer, H.~Stoecker and W.~Greiner,
  %``Quasiconfinement in the SU(3) gluon plasma,''
  Z.\ Phys.\ C {\bf 56}, 325 (1992).
  doi:10.1007/BF01555532
  
  \bibitem{Kapusta:1982qd} 
  J.~I.~Kapusta and K.~A.~Olive,
  %``Thermodynamics of Hadrons: Delimiting the Temperature,''
  Nucl.\ Phys.\ A {\bf 408}, 478 (1983).
  doi:10.1016/0375-9474(83)90241-5
  
  
  \bibitem{Singh:1991np} 
  C.~P.~Singh, B.~K.~Patra and K.~K.~Singh,
  %``Thermodynamically consistent EOS for hot dense hadron gas,''
  Phys.\ Lett.\ B {\bf 387}, 680 (1996).
  doi:10.1016/0370-2693(96)01117-3
  
  
  \bibitem{Karsch:2003vd} 
  F.~Karsch, K.~Redlich and A.~Tawfik,
  %``Hadron resonance mass spectrum and lattice QCD thermodynamics,''
  Eur.\ Phys.\ J.\ C {\bf 29}, 549 (2003)
  doi:10.1140/epjc/s2003-01228-y
  [hep-ph/0303108].
  
  \bibitem{Karsch:2003zq} 
  F.~Karsch, K.~Redlich and A.~Tawfik,
  %``Thermodynamics at nonzero baryon number density: A Comparison of lattice and hadron resonance gas model calculations,''
  Phys.\ Lett.\ B {\bf 571}, 67 (2003)
  doi:10.1016/j.physletb.2003.08.001
  [hep-ph/0306208].
  
  \bibitem{Redlich:2004gp} 
  K.~Redlich, F.~Karsch and A.~Tawfik,
  %``Heavy ion collisions and lattice QCD at finite baryon density,''
  J.\ Phys.\ G {\bf 30}, S1271 (2004)
  doi:10.1088/0954-3899/30/8/106
  [nucl-th/0404009].
  
  
  
  
  \bibitem{Albright:2014gva} 
  M.~Albright, J.~Kapusta and C.~Young,
  %``Matching Excluded Volume Hadron Resonance Gas Models and Perturbative QCD to Lattice Calculations,''
  Phys.\ Rev.\ C {\bf 90}, no. 2, 024915 (2014)
  doi:10.1103/PhysRevC.90.024915
  [arXiv:1404.7540 [nucl-th]].
  
  \bibitem{Andronic:2012ut} 
  A.~Andronic, P.~Braun-Munzinger, J.~Stachel and M.~Winn,
  %``Interacting hadron resonance gas meets lattice QCD,''
  Phys.\ Lett.\ B {\bf 718}, 80 (2012)
  doi:10.1016/j.physletb.2012.10.001
  [arXiv:1201.0693 [nucl-th]].
  
   \bibitem{Kadam:2014cua} 
  G.~P.~Kadam and H.~Mishra,
  %``Bulk and shear viscosities of hot and dense hadron gas,''
  Nucl.\ Phys.\ A {\bf 934}, 133 (2014)
  doi:10.1016/j.nuclphysa.2014.12.004
  [arXiv:1408.6329 [hep-ph]].
  
  
  \bibitem{Vovchenko:2014pka} 
  V.~Vovchenko, D.~V.~Anchishkin and M.~I.~Gorenstein,
  %``Hadron Resonance Gas Equation of State from Lattice QCD,''
  Phys.\ Rev.\ C {\bf 91}, no. 2, 024905 (2015)
  doi:10.1103/PhysRevC.91.024905
  [arXiv:1412.5478 [nucl-th]].
  
  
 

  

\bibitem{Gale:2013da} 
C.~Gale, S.~Jeon and B.~Schenke,
%``Hydrodynamic Modeling of Heavy-Ion Collisions,''
Int.\ J.\ Mod.\ Phys.\ A {\bf 28}, 1340011 (2013).
Phys.\ Rev.\ C {\bf 90}, no. 3, 034907 (2014).
  
 
\bibitem{Schenke:2011qd} 
B.~Schenke,
%``Flow in heavy-ion collisions - Theory Perspective,''
J.\ Phys.\ G {\bf 38}, 124009 (2011).

\bibitem{Shen:2012vn} 
C.~Shen and U.~Heinz,
%``Collision Energy Dependence of Viscous Hydrodynamic Flow in Relativistic Heavy-Ion Collisions,''
Phys.\ Rev.\ C {\bf 85}, 054902 (2012).

\bibitem{Kolb:2003dz} 
P.~F.~Kolb and U.~W.~Heinz,
%``Hydrodynamic description of ultrarelativistic heavy ion collisions,''
In *Hwa, R.C. (ed.) et al.: Quark gluon plasma* 634-714
[nucl-th/0305084].
\bibitem{Teaney:2000cw} 
D.~Teaney, J.~Lauret and E.~V.~Shuryak,
%``Flow at the SPS and RHIC as a quark gluon plasma signature,''
Phys.\ Rev.\ Lett.\  {\bf 86}, 4783 (2001).
\bibitem{DelZanna:2013eua} 
L.~Del Zanna {\it et al.},
%``Relativistic viscous hydrodynamics for heavy-ion collisions with ECHO-QGP,''
Eur.\ Phys.\ J.\ C {\bf 73}, 2524 (2013).
\bibitem{Karpenko:2013wva} 
I.~Karpenko, P.~Huovinen and M.~Bleicher,
%``A 3+1 dimensional viscous hydrodynamic code for relativistic heavy ion collisions,''
Comput.\ Phys.\ Commun.\  {\bf 185}, 3016 (2014).

\bibitem{Holopainen:2011hq} 
H.~Holopainen, H.~Niemi and K.~J.~Eskola,
%``Elliptic flow from event-by-event hydrodynamics,''
J.\ Phys.\ G {\bf 38}, 124164 (2011).
\bibitem{Jaiswal:2015mxa} 
A.~Jaiswal, B.~Friman and K.~Redlich,
%``Relativistic second-order dissipative hydrodynamics at finite chemical potential,''
Phys.\ Lett.\ B {\bf 751}, 548 (2015).

\bibitem{Xu:2004mz} 
Z.~Xu and C.~Greiner,
%``Thermalization of gluons in ultrarelativistic heavy ion collisions by including three-body interactions in a parton cascade,''
Phys.\ Rev.\ C {\bf 71}, 064901 (2005).

\bibitem{Bouras:2010hm} 
I.~Bouras, E.~Molnar, H.~Niemi, Z.~Xu, A.~El, O.~Fochler, C.~Greiner and D.~H.~Rischke,
%``Investigation of shock waves in the relativistic Riemann problem: A Comparison of viscous fluid dynamics to kinetic theory,''
Phys.\ Rev.\ C {\bf 82}, 024910 (2010).

\bibitem{Bouras:2012mh} 
I.~Bouras, A.~El, O.~Fochler, H.~Niemi, Z.~Xu and C.~Greiner,
%``Transition from ideal to viscous Mach cones in a kinetic transport approach,''
Phys.\ Lett.\ B {\bf 710}, 641 (2012).

\bibitem{Fochler:2010wn} 
O.~Fochler, Z.~Xu and C.~Greiner,
%``Energy loss in a partonic transport model including bremsstrahlung processes,''
Phys.\ Rev.\ C {\bf 82}, 024907 (2010).
\bibitem{Wesp:2011yy} 
C.~Wesp, A.~El, F.~Reining, Z.~Xu, I.~Bouras and C.~Greiner,
%``Calculation of shear viscosity using Green-Kubo relations within a parton cascade,''
Phys.\ Rev.\ C {\bf 84}, 054911 (2011).
\bibitem{Uphoff:2012gb} 
J.~Uphoff, O.~Fochler, Z.~Xu and C.~Greiner,
%``Open Heavy Flavor in Pb+Pb Collisions at $\sqrt{s}=2.76$ TeV within a Transport Model,''
Phys.\ Lett.\ B {\bf 717}, 430 (2012).
\bibitem{Greif:2013bb} 
M.~Greif, F.~Reining, I.~Bouras, G.~S.~Denicol, Z.~Xu and C.~Greiner,
%``Heat conductivity in relativistic systems investigated using a partonic cascade,''
Phys.\ Rev.\ E {\bf 87}, 033019 (2013).

\bibitem{Danielewicz:1984ww} 
P.~Danielewicz and M.~Gyulassy,
%``Dissipative Phenomena in Quark Gluon Plasmas,''
Phys.\ Rev.\ D {\bf 31}, 53 (1985).	

 \bibitem{Gyulassy:2004zy} 
M.~Gyulassy and L.~McLerran,
%``New forms of QCD matter discovered at RHIC,''
Nucl.\ Phys.\ A {\bf 750}, 30 (2005).
\bibitem{Csernai:2006zz} 
L.~P.~Csernai, J.~I.~Kapusta and L.~D.~McLerran,
%``On the Strongly-Interacting Low-Viscosity Matter Created in Relativistic Nuclear Collisions,''
Phys.\ Rev.\ Lett.\  {\bf 97}, 152303 (2006).
\bibitem{Kovtun:2004de} 
P.~Kovtun, D.~T.~Son and A.~O.~Starinets,
%``Viscosity in strongly interacting quantum field theories from black hole physics,''
Phys.\ Rev.\ Lett.\  {\bf 94}, 111601 (2005).

\bibitem{Gavin:1985ph} 
 S.~Gavin,
 %``Transport Coefficients In Ultrarelativistic Heavy Ion Collisions,''
 Nucl.\ Phys.\ A {\bf 435}, 826 (1985).
\bibitem{Prakash:1993bt} 
  M.~Prakash, M.~Prakash, R.~Venugopalan and G.~Welke,
  %``Nonequilibrium properties of hadronic mixtures,''
  Phys.\ Rept.\  {\bf 227}, 321 (1993).
\bibitem{Dobado:2003wr} 
  A.~Dobado and F.~J.~Llanes-Estrada,
  %``The Viscosity of meson matter,''
  Phys.\ Rev.\ D {\bf 69}, 116004 (2004).
  \bibitem{Chen:2007xe} 
     J.~W.~Chen, Y.~H.~Li, Y.~F.~Liu and E.~Nakano,
     %``QCD viscosity to entropy density ratio in the hadronic phase,''
     Phys.\ Rev.\ D {\bf 76}, 114011 (2007).
\bibitem{Itakura:2007mx} 
  K.~Itakura, O.~Morimatsu and H.~Otomo,
  %``Shear viscosity of a hadronic gas mixture,''
  Phys.\ Rev.\ D {\bf 77}, 014014 (2008). 
\bibitem{Dobado:2009ek} 
  A.~Dobado, F.~J.~Llanes-Estrada and J.~M.~Torres-Rincon,
  %``Minimum of eta/s and the phase transition of the Linear Sigma Model in the large-N limit,''
  Phys.\ Rev.\ D {\bf 80}, 114015 (2009). 
  
 \bibitem{Demir:2008tr} 
   N.~Demir and S.~A.~Bass,
   %``Shear-Viscosity to Entropy-Density Ratio of a Relativistic Hadron Gas,''
   Phys.\ Rev.\ Lett.\  {\bf 102}, 172302 (2009).
 \bibitem{Puglisi:2014pda} 
A.~Puglisi, S.~Plumari and V.~Greco,
%``Shear viscosity η to electric conductivity σ$_{el}$ ratio for the quark–gluon plasma,''
Phys.\ Lett.\ B {\bf 751}, 326 (2015).

\bibitem{Thakur:2017hfc} 
L.~Thakur, P.~K.~Srivastava, G.~P.~Kadam, M.~George and H.~Mishra,
%``Shear viscosity $\eta$ to electrical conductivity $\sigma_{el}$ ratio for an anisotropic QGP,''
Phys.\ Rev.\ D {\bf 95}, 096009 (2017).  

 \bibitem{Tawfik:2010mb} 
  A.~Tawfik and M.~Wahba,
  %``Bulk and Shear Viscosity in Hagedorn Fluid,''
  Annalen Phys.\  {\bf 522}, 849 (2010)
  doi:10.1002/andp.201000056
  [arXiv:1005.3946 [hep-ph]].
  
  \bibitem{Kadam:2018jaj} 
  G.~Kadam, S.~Pawar and H.~Mishra,
  %``Estimating transport coefficients of interacting pion gas with K-matrix cross sections,''
  J.\ Phys.\ G {\bf 46}, no. 1, 015102 (2019)
  doi:10.1088/1361-6471/aaeba2
  [arXiv:1807.05370 [nucl-th]].
  
  \bibitem{Denicol:2013nua} 
  G.~S.~Denicol, C.~Gale, S.~Jeon and J.~Noronha,
  %``Fluid behavior of a baryon-rich hadron resonance gas,''
  Phys.\ Rev.\ C {\bf 88}, no. 6, 064901 (2013)
  doi:10.1103/PhysRevC.88.064901
  [arXiv:1308.1923 [nucl-th]].
  
  \bibitem{Khvorostukhin:2010aj} 
  A.~S.~Khvorostukhin, V.~D.~Toneev and D.~N.~Voskresensky,
  %``Viscosity Coefficients for Hadron and Quark-Gluon Phases,''
  Nucl.\ Phys.\ A {\bf 845}, 106 (2010)
  doi:10.1016/j.nuclphysa.2010.05.058
  [arXiv:1003.3531 [nucl-th]].
  
 
  
  \bibitem{Kadam:2015xsa} 
  G.~P.~Kadam and H.~Mishra,
  %``Dissipative properties of hot and dense hadronic matter in an excluded-volume hadron resonance gas model,''
  Phys.\ Rev.\ C {\bf 92}, no. 3, 035203 (2015)
  doi:10.1103/PhysRevC.92.035203
  [arXiv:1506.04613 [hep-ph]].
  
  
  
  
  
  \bibitem{FernandezFraile:2009mi} 
  D.~Fernandez-Fraile and A.~Gomez Nicola,
  %``Transport coefficients and resonances for a meson gas in Chiral Perturbation Theory,''
  Eur.\ Phys.\ J.\ C {\bf 62}, 37 (2009)
  doi:10.1140/epjc/s10052-009-0935-0
  [arXiv:0902.4829 [hep-ph]].
  
  \bibitem{Kadam:2017iaz} 
  G.~P.~Kadam, H.~Mishra and L.~Thakur,
  %``Electrical and thermal conductivities of hot and dense hadronic matter,''
  Phys.\ Rev.\ D {\bf 98}, no. 11, 114001 (2018)
  doi:10.1103/PhysRevD.98.114001
  [arXiv:1712.03805 [hep-ph]].

  
 
  
  \bibitem{Liao:2009gb} 
  J.~Liao and V.~Koch,
  %``On the Fluidity and Super-Criticality of the QCD matter at RHIC,''
  Phys.\ Rev.\ C {\bf 81}, 014902 (2010)
  doi:10.1103/PhysRevC.81.014902
  [arXiv:0909.3105 [hep-ph]].
  

  
  \bibitem{Cleymans:2005xv} 
  J.~Cleymans, H.~Oeschler, K.~Redlich and S.~Wheaton,
  %``Comparison of chemical freeze-out criteria in heavy-ion collisions,''
  Phys.\ Rev.\ C {\bf 73}, 034905 (2006)
  doi:10.1103/PhysRevC.73.034905
  [hep-ph/0511094].
  
  \bibitem{Bugaev:2013jza} 
  K.~A.~Bugaev, A.~I.~Ivanytskyi, D.~R.~Oliinychenko, E.~G.~Nikonov, V.~V.~Sagun and G.~M.~Zinovjev,
  %``Non-smooth Chemical Freeze-out and Apparent Width of Wide Resonances and Quark Gluon Bags in a Thermal Environment,''
  Ukr.\ J.\ Phys.\  {\bf 60}, 181 (2015)
  [arXiv:1312.4367 [nucl-th]].
  
  \bibitem{Tawfik:2012si} 
  A.~Tawfik,
  %``On the Higher Moments of Particle Multiplicity, Chemical Freeze-Out and QCD Critical Endpoint,''
  Adv.\ High Energy Phys.\  {\bf 2013}, 574871 (2013)
  doi:10.1155/2013/574871
  [arXiv:1205.1761 [hep-ph]].
  
  \bibitem{Tawfik:2013dba} 
  A.~Tawfik,
  %``Chemical Freeze-Out and Higher Order Multiplicity Moments,''
  Nucl.\ Phys.\ A {\bf 922}, 225 (2014)
  doi:10.1016/j.nuclphysa.2013.12.008
  [arXiv:1306.1025 [hep-ph]].
  
  \bibitem{Tawfik:2013eua} 
  A.~Tawfik,
  %``Constant Trace Anomaly as a Universal Condition for the Chemical Freeze-Out,''
  Phys.\ Rev.\ C {\bf 88}, 035203 (2013)
  doi:10.1103/PhysRevC.88.035203
  [arXiv:1308.1712 [hep-ph]].
  
  \bibitem{Tawfik:2016sqd} 
  N.~A.~Tawfik, L.~I.~Abou-Salem, A.~G.~Shalaby, M.~Hanafy, A.~Sorin, O.~Rogachevsky and W.~Scheinast,
  %``Particle production and chemical freezeout from the hybrid UrQMD approach at NICA energies,''
  Eur.\ Phys.\ J.\ A {\bf 52}, no. 10, 324 (2016)
  doi:10.1140/epja/i2016-16324-6
  [arXiv:1609.08423 [nucl-th]].
  
  \bibitem{Tawfik:2018ahq} 
  A.~N.~Tawfik, H.~Yassin and E.~R.~Abo Elyazeed,
  %``Chemical freezeout parameters within generic nonextensive statistics,''
  Indian J.\ Phys.\  {\bf 92}, no. 10, 1325 (2018)
  doi:10.1007/s12648-018-1216-2
  [arXiv:1802.04912 [nucl-th]].
  
  
 
  
  
  
  
  
  
  

  
  \end{thebibliography}
\end{document}